\documentstyle[a4p,12pt,epsfig,subfigure,array,cite,pennames]{article}
\newcommand{\Date}      {\rightline 08-09-1998}
\newcommand{\PPEnum} {CERN-EP/98-???}
\newcommand{\smcap}[1] {\caption[]{\small #1}}
\newcommand{\inmath}[1] {\ifmmode#1\else$#1$\fi}
\newcommand{\definmath}[2] {\def#1{\ifmmode#2\else$#2$\fi}}
\definmath{\nctoe}{0.103}
\definmath{\estatctoe}{0.009}
\definmath{\esysctoe}{^{+0.009}_{-0.008}}
\definmath{\nctom}{0.090}
\definmath{\estatctom}{0.007}
\definmath{\esysctom}{^{+0.007}_{-0.006}}

\definmath{\nctol}{0.095}
\definmath{\estatctol}{0.006}
\definmath{\esysctol}{^{+0.007}_{-0.006}}
\definmath{\GeV}  {\mathrm{GeV}}
\definmath{\GeVc} {\mathrm{GeV}\!/c}
\definmath{\GeVcc}   {\mathrm{GeV}\!/c^2}
\definmath{\MeV}  {\mathrm{MeV}}
\definmath{\MeVc} {\mathrm{MeV}\!/c}
\definmath{\MeVcc}   {\mathrm{MeV}\!/c^2}
\definmath{\MVm}  {\mathrm{MV}\!/\mathrm{m}}
\definmath{\keV}  {\mathrm{keV}}
\definmath{\keVcm}   {\mathrm{keV}\!/\mathrm{cm}}
\definmath{\kV}      {\mathrm{kV}}
\definmath{\km}      {\mathrm{km}}
\definmath{\meter}   {\mathrm{m}}
\definmath{\cm}      {\mathrm{cm}}
\definmath{\mm}      {\mathrm{mm}}
\definmath{\micron}  {\mu\mathrm{m}}
\definmath{\nm}      {\mathrm{nm}}
\definmath{\kg}      {\mathrm{kg}}
\definmath{\gram} {\mathrm{g}}
\definmath{\second}  {\mathrm{s}}
\definmath{\microsec}   {\mu\mathrm{s}}
\definmath{\degree}  {^\circ}
\definmath{\degC} {^\circ\mathrm{C}}
\definmath{\ohm}  {\Omega}
\definmath{\Mohm} {\mathrm{M}\Omega}
\definmath{\rad}  {\mathrm{rad}}
\definmath{\mrad} {\mathrm{mrad}}
\definmath{\nb}      {\mathrm{nb}}
\definmath{\dEdx} {{\mathrm d}E/{\mathrm d}x}
\def\cc{\ifmmode {{\mathrm c\bar{\mathrm c}}}
    \else {${\mathrm c\bar{\mathrm c}}$} \fi}
\def\bb{\ifmmode {{\mathrm b\bar{\mathrm b}}}
    \else {${\mathrm b\bar{\mathrm b}}$} \fi}
\def\qq{\ifmmode {{\mathrm q\bar{\mathrm q}}}
    \else {${\mathrm q\bar{\mathrm q}}$} \fi}
\definmath{\PWpm} {\mathrm{W}^{\pm}}      % W+-
\definmath{\Pgtp} {\tau^{+}}        % tau+
\definmath{\Pgtm} {\tau^{-}}        % tau-
\definmath{\Pgtpm}   {\tau^{\pm}}         % tau+-
\definmath{\Pgn}  {\nu}          % neutrino
\definmath{\Pagn} {\overline{\nu}}     % anti-neutrino
\definmath{\Pq}      {\mathrm{q}}
\definmath{\Paq}  {\overline{\mathrm{q}}}
\definmath{\Pf}      {\mathrm{f}}
\definmath{\Paf}  {\overline{\mathrm{f}}}
\definmath{\Pu}      {\mathrm{u}}
\definmath{\Pau}  {\overline{\mathrm{u}}}
\definmath{\Pd}      {\mathrm{d}}
\definmath{\Pad}  {\overline{\mathrm{d}}}
\definmath{\Ps}      {\mathrm{s}}
\definmath{\Pas}  {\overline{\mathrm{s}}}
\definmath{\Pc}      {\mathrm{c}}
\definmath{\Pac}  {\overline{\mathrm{c}}}
\definmath{\Pb}      {\mathrm{b}}
\definmath{\Pab}  {\overline{\mathrm{b}}}
\definmath{\Pt}      {\mathrm{t}}
\definmath{\Pat}  {\overline{\mathrm{t}}}
\definmath{\Pap}  {\overline{\mathrm{p}}}
\definmath{\Pan}  {\overline{\mathrm{n}}}
\definmath{\PaD}  {\overline{\mathrm{D}}}
\definmath{\PaDz} {\overline{\mathrm{D}}^{0}}
\definmath{\PaB}  {\overline{\mathrm{B}}}
\definmath{\PaBz} {\overline{\mathrm{B}}^{0}}
\definmath{\PsDpm}   {\mathrm{D}^{\pm}_{\mathrm{s}}}  % Ds+-
\definmath{\PcgLpm}  {\Lambda^{\pm}_{\mathrm{c}}}  % Lambda_c+-
\definmath{\PcgL}  {\Lambda_{\mathrm{c}}}  % Lambda_c
\definmath{\PsBz}  {\overline{\mathrm{B}^0_{\mathrm{s}}}} % Bs0
\definmath{\PbgL}  {\Lambda_{\mathrm{c}}}  % Lambda_c
\definmath{\PD0}  {\mathrm{D}^{0}}     % D
\definmath{\PDp} {{\mathrm{D}^{+}}}  % D+-
\definmath{\PDz}  {\mathrm{D}^{0}}
\definmath{\PDs}  {{\mathrm{D}^{*}}}     % D*
\definmath{\PDsp} {{\mathrm{D}^{*+}}}     % D*+
\definmath{\PDsm} {{\mathrm{D}^{*-}}}     % D*-
\definmath{\PDspm} {{\mathrm{D}^{*\pm}}}     % D*+-
\definmath{\PK}{{\mathrm{K}^-}}
\definmath{\PKs}{{\mathrm{K}_{\rm S}}}
\definmath{\PPp}{{\pi^+}}
\definmath{\PPm}{{\pi^-}}
\definmath{\PEp}{{\rm e}^+}
\definmath{\PEm}{{\rm e}^-}
\definmath{\PE} {{\rm e}}
\definmath{\PMp}{{\mu }^+}
\definmath{\PMm}{{\mu }^-}
\definmath{\PM} {{\mu }}

\definmath{\PLp}{{\ell }^+}
\definmath{\PLm}{{\ell }^-}
\definmath{\PL}{{\ell }}

\definmath{\PgLz} {{\Lambda^{0}}}        % Lambda0
\def\D0{\ifmmode {{\mathrm D^0}} \else {${\mathrm D^0}$}\fi}
\def\Z0{\ifmmode {{\mathrm Z^0}} \else {${\mathrm Z^0}$}\fi}
\newcommand{\ccbar}  {\Pc\Pac}
\newcommand{\bbbar}  {\Pb\Pab}

\newcommand{\btol}      {\Pb\to\ell}

\newcommand{\btoctol}      {\Pb\to\Pc\to\ell}
\newcommand{\btocbartol}   {\Pb\to\Pac\to\ell}
\newcommand{\btotautol}    {\Pb\to\Pgt\to\ell}
\newcommand{\btoJPsitol}    {\Pb\to{\rm J}/\Psi\to\ell^+\ell^-}

\newcommand{\fbtol}      {{\rm B}(\Pb\to\ell)}

\newcommand{\fbtoctol}      {{\rm B}(\Pb\to\Pc\to\ell)}
\newcommand{\fbtocbartol}   {{\rm B}(\Pb\to\Pac\to\ell)}
\newcommand{\fbtotautol}    {{\rm B}(\Pb\to\Pgt\to\ell)}
\newcommand{\fbtoJPsitol}    {{\rm B}(\Pb\to{\rm J}/\Psi\to\ell^+\ell^-)}

\newcommand{\ctol}      {\Pc\to\ell}

\newcommand{\ctom}      {\Pc\to\mu}

\newcommand{\ctoe}      {\Pc\to{\rm e}}

\newcommand{\fctol}      {{\rm B}(\Pc\to\ell)}

\newcommand{\fctom}      {{\rm B}(\Pc\to\mu)}

\newcommand{\fctoe}      {{\rm B}(\Pc\to{\rm e})}

\definmath{\BR}{\mathrm B}
\definmath{\NRQ}{N_{\PDsp,\ell^-}}
\definmath{\NWQ}{N_{\PDsp,\ell^+}}
\definmath{\dbtagWQ}{N_{\PDsp,\ell^+}}
\definmath{\NDLc}{{N_{\PDsp,\ell^-}^{\Pc}}}
\definmath{\fc}{f_{\Pc}^{\PDsp}}

\newcommand {\downto}
         {\mbox{ \begin{picture}(14,10)
                    \put(0,10){\line(0,-1){5.0}}
                    \put(2,5){\oval(4,4)[bl]}
                    \put(1,0){\makebox(0,0)[bl]{$\rightarrow$}}
                 \end{picture} }}

\def\etal{et al.}

\begin{document}
% \input{/u/meyer/latex/tiesmath.tex}
% \psdraft
%%%%%%%%%%%%%%%%%%%%%%%%%%%%%%%%%%%%%%%%%%%%%%%%%%%%%%%%%%%%%%%%%%%%%%%%
%
%  Title Page
%
\begin{titlepage}
%
%     Header
%
\begin{center}
   \Large
    EUROPEAN LABORATORY FOR PARTICLE PHYSICS
\end{center}
\bigskip
\rightline{\PPEnum}
\rightline{\Date}
\vskip 2cm
%
%     Main title
%
\begin{center}
    \huge\bf\boldmath
Measurement of the Semileptonic Branching Ratio of 
Charm Hadrons Produced in {$\boldmath \PZz\to\ccbar$} Decays
\end{center}
%\vspace{1cm}
\bigskip
%
%     Author names
%
\begin{center}
{\bf \large THE OPAL COLLABORATION}\\
\end{center}
\vskip 1cm

\bigskip
%
%     Abstract
%
\begin{abstract}%=======================================================

\noindent 
The inclusive charm hadron semileptonic branching fractions
$\fctoe$ and $\fctom$ in $\PZz\to\ccbar$ events have been 
determined using  $4.4$ million hadronic $\PZz$ decays collected
with the OPAL detector at LEP. 
A charm-enriched sample is obtained by selecting events
with reconstructed $\PDspm$ mesons.
Using leptons found in the hemisphere opposite that of the $\PDspm$ mesons,
the semileptonic branching fractions of charm hadrons are measured to be
$$ \fctoe = \nctoe \pm \estatctoe \; \esysctoe  ~~~{\rm and} ~~~ 
   \fctom = \nctom \pm \estatctom \; \esysctom\ ,$$
where the first errors are in each case  statistical and the second systematic.
Combining these measurements, an inclusive semileptonic branching 
fraction of charm hadrons of 
$$ \fctol = \nctol \pm \estatctol \; \esysctol $$
is obtained. 

\end{abstract} 
\vspace{2cm}
%\begin{center}
%This is a final dispatch. Comments to 
%Ties.Behnke{@}desy.de or meyer@bolte.desy.de, \\
%till Friday, 4/9/1998, 12:00, please.
%\end{center}
\begin{center}
Submitted to European Journal of Physics
\end{center}
\end{titlepage}

\newpage
\begin{center}{
%begin authorlist PLEASE DO NOT DELETE THIS COMMENT
G.\thinspace Abbiendi$^{  2}$,
K.\thinspace Ackerstaff$^{  8}$,
G.\thinspace Alexander$^{ 23}$,
J.\thinspace Allison$^{ 16}$,
N.\thinspace Altekamp$^{  5}$,
K.J.\thinspace Anderson$^{  9}$,
S.\thinspace Anderson$^{ 12}$,
S.\thinspace Arcelli$^{ 17}$,
S.\thinspace Asai$^{ 24}$,
S.F.\thinspace Ashby$^{  1}$
% ...
%end authorlist PLEASE DO NOT DELETE THIS COMMENT
}\end{center}\bigskip
\bigskip
%begin institutes
$^{  1}$School of Physics and Astronomy, University of Birmingham,
Birmingham B15 2TT, UK
\newline
$^{  2}$Dipartimento di Fisica dell' Universit{\`a} di Bologna and INFN,
I-40126 Bologna, Italy
\newline
$^{  3}$Physikalisches Institut, Universit{\"a}t Bonn,
D-53115 Bonn, Germany
\newline
$^{  4}$Department of Physics, University of California,
Riverside CA 92521, USA
\newline
$^{  5}$Cavendish Laboratory, Cambridge CB3 0HE, UK
\newline
$^{  6}$Ottawa-Carleton Institute for Physics,
Department of Physics, Carleton University,
Ottawa, Ontario K1S 5B6, Canada
\newline
$^{  7}$Centre for Research in Particle Physics,
Carleton University, Ottawa, Ontario K1S 5B6, Canada
\newline
$^{  8}$CERN, European Organisation for Particle Physics,
CH-1211 Geneva 23, Switzerland
\newline
$^{  9}$Enrico Fermi Institute and Department of Physics,
University of Chicago, Chicago IL 60637, USA
\newline
$^{ 10}$Fakult{\"a}t f{\"u}r Physik, Albert Ludwigs Universit{\"a}t,
D-79104 Freiburg, Germany
\newline
$^{ 11}$Physikalisches Institut, Universit{\"a}t
Heidelberg, D-69120 Heidelberg, Germany
\newline
$^{ 12}$Indiana University, Department of Physics,
Swain Hall West 117, Bloomington IN 47405, USA
\newline
$^{ 13}$Queen Mary and Westfield College, University of London,
London E1 4NS, UK
\newline
$^{ 14}$Technische Hochschule Aachen, III Physikalisches Institut,
Sommerfeldstrasse 26-28, D-52056 Aachen, Germany
\newline
$^{ 15}$University College London, London WC1E 6BT, UK
\newline
$^{ 16}$Department of Physics, Schuster Laboratory, The University,
Manchester M13 9PL, UK
\newline
$^{ 17}$Department of Physics, University of Maryland,
College Park, MD 20742, USA
\newline
$^{ 18}$Laboratoire de Physique Nucl{\'e}aire, Universit{\'e} de Montr{\'e}al,
Montr{\'e}al, Quebec H3C 3J7, Canada
\newline
$^{ 19}$University of Oregon, Department of Physics, Eugene
OR 97403, USA
\newline
$^{ 20}$CLRC Rutherford Appleton Laboratory, Chilton,
Didcot, Oxfordshire OX11 0QX, UK
\newline
$^{ 22}$Department of Physics, Technion-Israel Institute of
Technology, Haifa 32000, Israel
\newline
$^{ 23}$Department of Physics and Astronomy, Tel Aviv University,
Tel Aviv 69978, Israel
\newline
$^{ 24}$International Centre for Elementary Particle Physics and
Department of Physics, University of Tokyo, Tokyo 113, and
Kobe University, Kobe 657, Japan
\newline
$^{ 25}$Institute of Physical and Environmental Sciences,
Brunel University, Uxbridge, Middlesex UB8 3PH, UK
\newline
$^{ 26}$Particle Physics Department, Weizmann Institute of Science,
Rehovot 76100, Israel
\newline
$^{ 27}$Universit{\"a}t Hamburg/DESY, II Institut f{\"u}r Experimental
Physik, Notkestrasse 85, D-22607 Hamburg, Germany
\newline
$^{ 28}$University of Victoria, Department of Physics, P O Box 3055,
Victoria BC V8W 3P6, Canada
\newline
$^{ 29}$University of British Columbia, Department of Physics,
Vancouver BC V6T 1Z1, Canada
\newline
$^{ 30}$University of Alberta,  Department of Physics,
Edmonton AB T6G 2J1, Canada
\newline
$^{ 31}$Duke University, Dept of Physics,
Durham, NC 27708-0305, USA
\newline
$^{ 32}$Research Institute for Particle and Nuclear Physics,
H-1525 Budapest, P O  Box 49, Hungary
\newline
$^{ 33}$Institute of Nuclear Research,
H-4001 Debrecen, P O  Box 51, Hungary
\newline
$^{ 34}$Ludwigs-Maximilians-Universit{\"a}t M{\"u}nchen,
Sektion Physik, Am Coulombwall 1, D-85748 Garching, Germany
\newline
%end institutes
\bigskip\newline
%begin notes
$^{  a}$ and at TRIUMF, Vancouver, Canada V6T 2A3
\newline
$^{  b}$ and Royal Society University Research Fellow
\newline
$^{  c}$ and Institute of Nuclear Research, Debrecen, Hungary
\newline
$^{  d}$ and Department of Experimental Physics, Lajos Kossuth
University, Debrecen, Hungary
\newline
$^{  e}$ on leave of absence from the University of Freiburg
\newline
\newpage
%%%%%%%%%%%%%%%%%%%%%%%%%%%%%%%%%%%%%%%%%%%%%%%%%%%%%%%%%%%%%%%%%%%%%%%%

\section{Introduction}
The inclusive charm hadron semileptonic branching fractions
$\fctoe$  and $\fctom$ are defined as the 
average of the semileptonic branching ratios of 
weakly decaying charm hadrons weighted by their production 
rates in prompt charm events, $\PZz\to\ccbar$. 
%This branching fraction is of general theoretical interest, since 
%the semileptonic decays of heavy quarks are one of 
%the clearest sources of information about heavy quarks. 
%For charm quarks in particular the unexpectedly large differences between 
%the lifetimes of different charmed mesons is believed to 
%be related to properties of hadronic decays, which can be 
%studied most easily in semileptonic decays. 
%The inclusive semileptonic branching ratio of charm hadrons 
Inclusive semileptonic branching ratios are a 
means to investigate the dynamics of heavy quark decays, 
and have been studied in much detail for bottom quarks~\cite{bib-PDG}.
% at LEP and elsewhere .
The inclusive semileptonic branching 
ratio of charm hadrons 
has not previously been measured at LEP, even though it 
is an important input to a number of  
measurements performed at energies around the $\PZz$ resonance~\cite{bib-moenig}.
%  which makes it desirable to have LEP 
% results on this quantity. 

The inclusive semileptonic branching fraction of charm hadrons 
has so far been measured at centre-of-mass 
energies significantly below the 
$\PZz$ mass~\cite{bib-cl1,bib-cl2}. 
Many of these measurements depend strongly on the modelling 
of the $\Pb\to\ell$ background in the sample. In this paper 
a measurement of $\fctoe$ and $\fctom$  is presented 
which is much less dependent on the bottom background, since
it is done in a sample of events enriched in $\PZz\to\ccbar$ decays.
This sample is prepared by selecting highly energetic 
$\PDsp$ mesons\footnote{Throughout this note charge 
conjugation is always implied, unless explicitly stated otherwise.}.
%in one hemisphere of the event. 
The hemisphere opposite to the one containing the $\PDsp$ meson 
is searched for a lepton, yielding a measurement of the 
inclusive semileptonic branching fraction of charm hadrons.

The paper  is organised as follows. The principle of the analysis,
in particular the method used to subtract the background, is 
discussed in section \ref{sec-analysis}. 
After a brief review of the  event selection in section \ref{sec-hadsel}, the identification 
of charm events using reconstructed $\PDsp$ mesons is described 
and the determination of the charm fraction in the sample is summarised
in section \ref{sec-charmtagging}.
The preparation of the lepton sample  in charm-tagged events and 
the measurement of the background in this sample is 
described in section~\ref{sec-leptag}, 
followed by the presentation of the results in section~\ref{sec-results}. 
Systematic errors are given in section~\ref{sec-sys}. 

\vspace{1cm}
\section{Analysis Principle}
\label{sec-analysis}
A sample of $\PZz\to\ccbar$ enriched events is found using  
reconstructed $\PDsp$ mesons. Each event is divided into two hemispheres by 
the plane perpendicular to the thrust axis.
% including the $\epem$ interaction point in the definition of the plane. 
Leptons are searched for in the hemisphere opposite the 
$\PDsp$ meson. 
Background is suppressed by requiring that the 
$\PDs$ and the $\ell$ have opposite charge.
The number of leptons found in the hemisphere opposite that of 
the $\PDs$ meson has contributions from prompt 
charm decays, $\Pc\to\ell$, from prompt bottom decays, $\Pb\to\ell$, 
from cascade decays, $\Pb\to\Pc\to\ell$, and from 
background. It can be written as 
\begin{eqnarray}
\label{eq-RQ}
\NRQ&=&N_{\PDsp} \cdot \Bigl \{
         \fc\; {\fctol}\;\epsilon_{\ell}^{\ctol} + \nonumber \\
&&(1-\fc)\; \Bigl[ 
\chi_{\mathrm eff}\; {\fbtol}\;\epsilon_{\ell}^{\btol} +
(1-\chi_{\rm eff})\; {\fbtoctol}\; \epsilon_{\ell}^{\btoctol} \Bigr] \Bigr\}+ 
N_{\rm bgd}^{+-}\ .
\end{eqnarray}
Here $N_{\PDsp}$ is the number of $\PDsp$ mesons found in the data sample, 
$\fc$ is the fraction of these $\PDsp$ mesons coming from 
$\PZz\to\ccbar$ events,
and $N_{\rm bgd}^{+-}$ is the number of background 
events, where either a 
$\PDsp$, a lepton or both are misidentified, but where the charge
correlation is correct between the two hemispheres. 
This background will be denoted as ``combinatorial background''.
The parameter $\chi_{\rm eff}$ is the effective mixing parameter 
for the mixture of neutral B mesons selected, and 
$\epsilon_{\ell}^{\ctol}, \epsilon_{\ell}^{\btol}$ and 
$\epsilon_{\ell}^{\btoctol}$ are the efficiencies to 
find a lepton opposite a $\PDsp$ in the channel indicated, with 
the correct charge correlation. To simplify this and the following equations, 
leptons produced in $\btocbartol$ decays and $\btotautol$ decays
are included in the $\btol$ decays. 
Since a pair of leptons, one with the correct 
and one with the wrong charge correlation is produced in  $\btoJPsitol$ decays they are 
equally split between the $\btol$ and the $\btoctol$ decays.

The goal of this analysis is the measurement of $\fctol$. It is 
extracted from $\NDLc$, the number of $\PZz\to\ccbar$ events where  
simultaneously a $\PDsp$ meson in one hemisphere 
and a lepton in the opposite hemisphere is found:
\begin{equation}
\label{eq-goal}
  \NDLc = {N_{\PDsp}}\; {\it{\fc}\;{\fctol}\; \epsilon_{\ell}^{\ctol}}\ .
\end{equation} 
A sample of events which does not contain contributions 
from prompt charm decays is prepared by selecting
events where the $\PDsp$ and the lepton have 
equal charge:
%This number is derived from eq.~\ref{eq-RQ} by subtracting the 
%contributions which do not originate from prompt charm decays 
%using a sample of events, where both the $\PDsp$ and the lepton 
%have the same charge: 
\begin{equation}
\label{eq-WQ}
\NWQ = N_{\PDsp}
(1-\fc)\;\left\{ (1-\chi_{\mathrm eff})\; {\fbtol}\;\epsilon_{\ell}^{\btol} +
\chi_{\rm eff}\; {\fbtoctol}\; \epsilon_{\ell}^{\btoctol}\right\} + 
N_{\rm bgd}^{++}\ .
\end{equation}
Here $N_{\rm bgd}^{++}$ is the number of combinatorial
background events in this wrong sign sample.
%The difference between this sample of like sign, $\PDsp\ell^+$, 
%candidates (eq.~\ref{eq-WQ})
%and the sample of oppositely charged candidates (eq.~\ref{eq-RQ})
%can be expressed as 
%the number of leptons opposite a $\PDsp$ meson in $\PZz\to\ccbar$ 
%events, $\NDLc$, plus some contributions from bottom and 
%from background events, $\Delta N_{\Pb}$ and $\Delta N_{\rm bgd}$. 
The number of leptons from charm hadron decays can be 
calculated by solving the two equations~\ref{eq-RQ} and \ref{eq-WQ} 
for $\NDLc$ defined in equation~\ref{eq-goal}. The solution can be written in 
terms of the difference of the two samples of events and 
two small corrections,
%
%Subtracting the number of wrong sign $\PDsp\ell^+$ candidates (eq.~\ref{eq-WQ})
%from the number of right sign  $\PDsp\ell^-$ candidates gives $\NDLc$, the number of leptons 
%opposite a $\PDsp$ meson in $\PZz\to\ccbar$ events. 
\begin{equation}
\label{eq-ctol}
\NDLc 
%N_{\PDsp} \fc \fctol \epsilon_{\ell}^{\ctol}
       = ( \NRQ - \NWQ ) - \Delta N_{\Pb} - \Delta N_{\rm bgd}\ .
\end{equation}
The first correction, $\Delta N_{\Pb}$, can be derived directly 
from equation~\ref{eq-RQ} and equation~\ref{eq-WQ} and reflects the fact 
that mixing affects both samples differently. It is 
calculable from the known branching ratios and 
the mixing parameter: 
\begin{equation}
\label{eq-B}
\Delta N_{\Pb} = N_{\PDsp}\;(1-\fc)\;(1-2 \chi_{\rm eff}) \; 
  \left \{ {\fbtoctol} \; \epsilon_{\ell}^{\btoctol} -
           {\fbtol}\; \epsilon_{\ell}^{\btol} \right \}\ .
\end{equation}
The second correction, $\Delta N_{\rm bgd}$, is the difference between
the combinatorial background term in both samples, 
$\Delta N_{\rm bgd} = N_{\rm bgd}^{+-} - N_{\rm bgd}^{++}$. 
This number is determined using both  data and Monte 
Carlo simulations, as will be discussed in section \ref{sec-bgd}. 
%The actual signal for this analysis, $\NDLc$, 
%therefore is given by:
Finally the inclusive semileptonic branching ratio of charm hadrons, 
$\fctol$, is calculated from equation~\ref{eq-goal} as 
\begin{equation}
  \fctol = {\NDLc} { 1 \over {N_{\PDsp}} {\it{\fc} \epsilon_{\ell}^{\ctol}}}\ ,
\end{equation} 
where the number of events with a $\PDsp$ meson, the number of leptons 
found in this sample, the efficiencies to reconstruct the 
leptons in the tagged charm sample, and 
$\Delta N_{\Pb}$ and $\Delta N_{\rm bgd}$ have to be known.
Each of these inputs will be discussed in the 
following sections. 

\section{The OPAL Detector and Event Selection} 
\label{sec-hadsel}
A detailed description of the OPAL detector can be found 
elsewhere \cite{bib-OPALdetector}. The most relevant parts of the 
detector for this analysis are the tracking chambers, the
electromagnetic calorimeter, and the muon chambers. The central 
detector provides precise measurements of the momenta of charged particles
by the curvature of their trajectories 
in a solenoidal magnetic field of $0.435\;$T. The electromagnetic calorimeter
consists
of approximately 12000 lead glass blocks, which completely cover 
the azimuthal range up to polar angles\footnote{The OPAL coordinate 
system is defined as a Cartesian coordinate system, with the 
$x$-axis pointing horizontally towards the 
centre of the LEP ring, the $z$-axis in the direction of the 
outgoing electrons, and the $y$-axis points approximately vertically upwards. 
The polar angle is measured with respect to the $z$-axis.} 
of $|\cos \theta|<0.98$. Nearly
the entire detector is surrounded with at least three layers of 
muon chambers, which are placed behind an approximately one meter thick iron
magnet flux  return yoke.

Hadronic $\PZz$ decays are selected using  the 
number of reconstructed charged 
tracks and the energy deposited in the calorimeter, as described 
in \cite{bib-OPALmh}. 
The analysis uses an initial sample 
of $4.4$ million hadronic  decays of the $\PZz$ collected between 1990 and 1995. 

Hadronic decays of the $\PZz$ have been simulated using the  JETSET 7.4 Monte 
Carlo model  \cite{bib-JETSET} with parameters tuned to 
the data \cite{bib-OPALtune}. The Monte Carlo samples are about five times larger than the 
collected data sample. 
Heavy quark fragmentation has been implemented using the 
model of Peterson \etal~ \cite{bib-PETERSON} with fragmentation parameters
determined from LEP data \cite{bib-LEPNIM}. 
The samples have been  passed through a detailed simulation of the OPAL 
detector \cite{bib-OPALGOPAL} before being analysed using the same 
programs as for the data. 
Jets are reconstructed in the events by the JADE jet finder using the E0 scheme with a 
cut-off parameter $x_{min}=49~\GeV^2$ \cite{bib-JADE}.

\section{Charm Tagging}
\label{sec-charmtagging}
The tagging of $\PZz\to\ccbar$ events is based on the 
reconstruction of charged $\PDsp$ mesons in five
different decay channels. 
The identification algorithm and the method to separate the 
different sources contributing to the observed $\PDsp$ signal 
have been presented in a previous OPAL paper~\cite{bib-ctodstar}, 
and will only be briefly reviewed. 

The $\PDsp$ mesons are reconstructed in the following
five %$\PDz$ 
decay channels:
\begin{center}
\begin{tabbing}
  \hspace{5cm} \= \hspace{5cm} \= \kill
  \> $\PDsp \rightarrow \PDz\pi^+$ \\
  \> $\phantom{\PDsp \rightarrow }\hspace{4pt} 
            \downto {\rm K^-}\pi^+$\ ,                  \> ``3-prong'' \\
  \> $\phantom{\PDsp \rightarrow }\hspace{4pt} 
            \downto {\rm K^-}{\rm e}^+ \nu_{{\rm e}}$\ ,\>``electron''\\
  \> $\phantom{\PDsp \rightarrow }\hspace{4pt} 
            \downto {\rm K^-}\mu^+\nu_{\mu}$\ ,          \>``muon''\\
  \> $\phantom{\PDsp \rightarrow }\hspace{4pt} 
            \downto {\rm K^-}\pi^+\Pgpz$\ ,             \> ``satellite'' \\
  \> $\phantom{\PDsp \rightarrow }\hspace{4pt} 
            \downto {\rm K^-}\pi^+\pi^-\pi^+$\ ,        \>``5-prong''\ .

\end{tabbing}
\end{center} 
The muon and the electron channels are collectively referred 
to as ``semileptonic''.
No attempt is made to reconstruct the $\Pgpz$ in the satellite channel, nor 
the neutrino in the two semileptonic channels. Electrons are identified based 
on their specific energy loss, $\dEdx$, in the central tracking chamber 
and the energy deposition in the 
electromagnetic calorimeter. An artificial neural network trained on 
simulated events is used 
to perform the selection \cite{bib-OPALANN}. 
Electrons from photon conversions are rejected as in \cite{bib-OPALleptonID}. 
Muons are selected using matching of track segments of the central tracking chambers 
and the muon chambers, as described in \cite{bib-OPALleptonID}. 
%Muon candidates are identified by associating tracks found in the central  
%tracking system with tracks in the outer  muon chambers \cite{bib-OPALleptonID}.
%No momentum cut is applied since the knowledge  of the lepton purity is not 
%required in the subsequent analysis. 
The purity of kaons is enhanced by requiring
that the $\dEdx$ measurement of the candidate is compatible with 
that expected for a kaon. If the track combination has an invariant 
mass $M_{\PDz}$ within the limits given in table \ref{tab-dstarcuts},
the combination is accepted as a $\PDz$ candidate.
The combinatorial background is reduced by a cut on   
the cosine of the helicity angle, $\cos \theta^*$, measured 
between the direction of the $\PDz$ in the laboratory frame and 
the direction of the kaon in the rest-frame of the $\PDz$ candidate.
Background is expected to peak at $\pm 1$ in this variable, while 
true $\PDz$ mesons are uniformly distributed. 
These $\PDz$ candidates are combined with a candidate for 
the pion in the $\PDsp\to\PDz\pi^+$ decay. Background from bottom decays and 
combinatorial background is reduced by selecting candidates 
with a large scaled energy, 
$x_{\PDsp}=E_{\PDsp}^{\rm cand}/E_{\rm beam}$. 
The final selection is made on the mass difference 
$\Delta M = M_{\PDsp}-M_{\PDz}$ between 
the $\PDsp$ candidate and the corresponding $\PDz$ candidate.
% A summary of the applied cuts for  each channel is shown in table \ref{tab-dstarcuts}. 

\begin{figure}[p]
\begin{center}
\mbox{\epsfig{figure=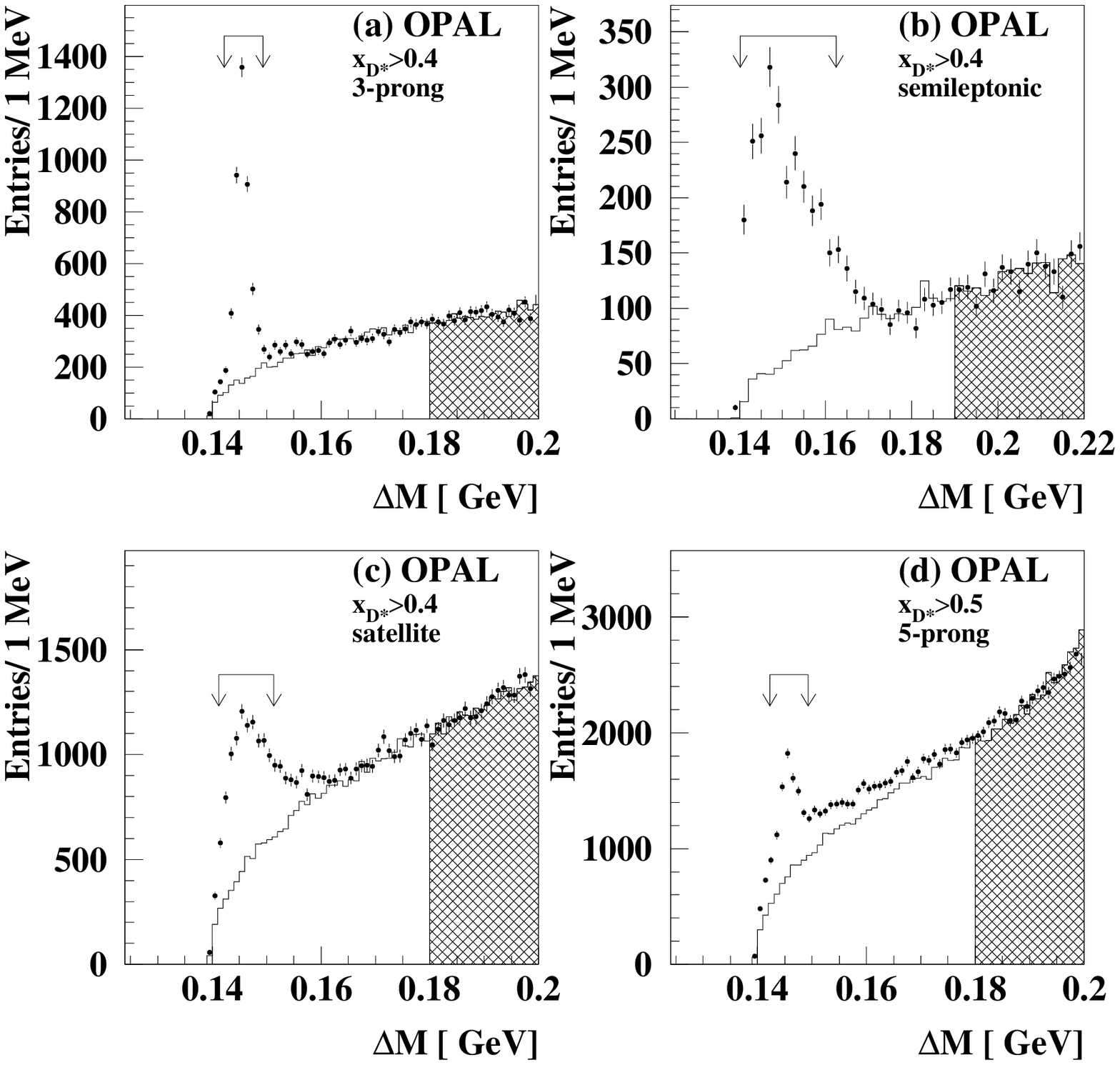,height=15cm,bbllx=3.5cm,bblly=5cm,bburx=16.5cm,bbury=22cm}}
\end{center}
\smcap{\label{fig-delm} Distributions of the mass difference 
$\Delta M = M_{\PDsp}-M_{\PDz}$
reconstructed in the four different $\PDsp$ channels. 
% Both semileptonic decays have been combined into one plot. 
The arrows indicate 
the range in $\Delta M$ considered as signal.  
The background estimator distributions are superimposed, 
normalised to the signal distribution at large values
of $\Delta M$ indicated by the cross-hatched area. 
Note that the significant tails in the $\Delta M$ distribution 
above the expected signal, particularly in (c) and (d), 
are caused by partially reconstructed 
$\PDsp$ mesons, and is properly treated by the background estimator
(see text).}
\end{figure}

If more than one $\PDsp$ candidate is found in an event, 
only one candidate is accepted according to the following procedure.
A 3-prong decay is 
preferred over a  semileptonic one, which in turn is preferred over a satellite, 
and a 5-prong decay is selected last. 
If more than one  candidate is found within the same decay channel, the 
one with $M_{\PDz}$ closest to its nominal value of $1.864~\GeV$ \cite{bib-PDG}
($1.6  ~\GeV$ for the satellite) is selected. 
In figure \ref{fig-delm}, the mass difference distributions 
$\Delta M = M_{\PDsp} - M_{\D0}$ are shown for the different channels. 
In total $27662$ candidates are selected in all five channels. 
\begin{table}[t]
\begin{center}
\begin{tabular}{|lc|c|c|c|c|}
\hline \multicolumn{2}{|c|}{}  & \multicolumn{4}{c|}{$\PD0$ decay mode}  \\
\hline
\multicolumn{2}{|c|}{ Variable} &      3-prong   & semileptonic  & satellite   & 5-prong   \\
\hline
\hline
\multicolumn{2}{|c|}{$x_{\PDsp}$ range  \rule[-1.7mm]{0mm}{6mm}}
                 &      $0.4$-$1.0$     & $0.4$-$1.0$     & $0.4$-$1.0$  &  
                    $0.5$-$1.0$        \\
\hline
\multicolumn{2}{|c|}{$M_{\PDz}$ [~\GeV] \rule[-1.7mm]{0mm}{6mm}}
                  &{$1.79$-$1.94$}  & $1.20$-$1.80$   & $1.41$-$1.77$     & 
                    $1.79$-$1.94$       \\
\hline
\multicolumn{2}{|c|}{$\Delta M$ [~\GeV]  \rule[-1.7mm]{0mm}{6mm}}
                  &{$0.142$-$0.149$}   & $0.140$-$0.162$ & $0.141$-$0.151$   & 
                    $0.142$-$0.149$        \\
\hline
$\cos \theta^*$ &$x_{\PDsp}<0.5$  \rule[-1.7mm]{0mm}{6mm}
                  & \multicolumn{3}{c|}{$-0.8 $-$ 0.8$}  &$-$
       \\ 
\hline
$\cos \theta^*$ &$x_{\PDsp}>0.5$  \rule[-1.7mm]{0mm}{6mm}
                  & \multicolumn{4}{c|}{$-0.9$-$1.0$} \\
\hline
$W_{\dEdx}^{\rm KK}$ \rule[-1.7mm]{0mm}{6mm} &
    $x_{\PDsp}<0.5$  & \multicolumn{3}{c|}{$ >0.1$}&$-$\\
\hline
\hline
\multicolumn{2}{|c|}{Relative abundance} &  0.231 & 0.121 & 0.355 & 0.293 \\
\hline
\multicolumn{2}{|c|}{Signal/background}  & 3.496 & 3.233 & 1.223 & 0.879 \\
\hline
\end{tabular}
\end{center}
\smcap{\label{tab-dstarcuts}
List of selection cuts used in the \protect$\PDsp$ reconstructions.
\protect$W_{\dEdx}^{\rm KK}$ is the probability
that the measured \protect$\dEdx$ value is compatible with that 
expected for a kaon at the measured momentum. This 
cut is only applied to the kaon candidate in the \protect$\PDz$ decay. 
The background distribution thus obtained is 
normalised to the candidate \protect$\Delta M$ distribution in the range 
\protect$0.18\;\GeV <\Delta M<0.20 \; \GeV$ ($0.19\;\GeV<\Delta M<0.22\;\GeV$ in the 
semileptonic channels).
%\protect$x_{\PDsp}=E_{\PDsp}^{\mathrm cand}/E_{\rm beam}$ is the scaled energy of the
%\protect$\PDsp$ candidate. \protect$M_{\PDz}$ is the mass of the \protect$\PDz$ candidate
%and \protect$\Delta M$ is the mass difference between the \protect$\PDz$ and the corresponding
%\protect$\PDsp$ candidate. \protect$\cos \theta^*$  is the helicity angle measured  between 
%the direction of the $\PDz$ candidate in the laboratory frame and the 
%direction of the kaon in the rest frame of the $\PDz$ candidate. 
In the last two lines of the 
table, the relative abundance of each channel and 
the signal/background ratio is given, as measured from the data.}
\end{table}

The selected sample of $\PDsp$ candidates has contributions from:
$\PDsp$ mesons produced in $\PZz\to\ccbar$ events (signal);
$\PDsp$ mesons produced in $\PZz\to\bbbar$ events; 
$\PDsp$ mesons produced in events where 
a $\ccbar$ pair is produced in the splitting of a gluon;
combinatorial background. 
The combinatorial background in the sample of $\PDsp$ mesons 
is subtracted on a statistical 
basis using an independent sample of background candidate 
events, selected based on a hemisphere mixing technique
first introduced in \cite{bib-OPALD*AFB}. 
The candidate for the pion in the $\PDsp\to\PDz\pi^+$ decay
is selected in the hemisphere opposite 
to the rest of the candidate, and reflected through the origin.
This sample of candidates has been shown to be 
an unbiased estimator of the combinatorial
background~\cite{bib-OPALD*AFB,bib-OPALD*AFB2}
and to reliably model the shape of the background. 
The contribution from gluon splitting is estimated and 
subtracted from the sample based 
on the OPAL measurement of the multiplicity of such events 
in hadronic $\PZz$ decays~\cite{bib-OPALgcc}. For 
the cuts used in this analysis, ${\rm g}\to\ccbar$ events 
contribute $(1.1 \pm 0.4)\%$ to the signal. 
After all corrections, and after combinatorial background subtraction,
$(15784 \pm 99)$ $\PDsp$ mesons are used in the subsequent  analysis. The error quoted is the statistical
uncertainty of the combinatorial background subtraction.  

% \subsection{Flavour Composition of the  \boldmath $\PDsp$ Sample}
\label{sec-flavsep} 
The remaining two sources of $\PDsp$ production, $\PZz\to\ccbar\to\PDsp X$ 
and $\PZz\to\bbbar\to\PDsp X$, are 
separated by applying three different flavour tagging
methods,
based on lifetime information, jet shapes and hemisphere charge
information,
as described in~\cite{bib-ctodstar}. 
Combining all $\PDsp$ channels,
the overall charm fraction is determined to be:
\begin{equation}
  \fc = 0.774 \pm 0.008 \pm 0.022,  
\end{equation}
where the first error is statistical and the second systematic. 
The dominant systematic 
errors are from the estimation of the background in the $\PDsp$ sample,
and from modelling the charm physics parameters used in the 
flavour separation. 
A breakdown of the systematic error into its components which are 
relevant for this analysis is given in table~\ref{tab-dstar}.
The errors are split into two groups: 
one group which is uncorrelated to errors encountered
when identifying leptons in this sample of events, 
and a second group of correlated errors.
In the latter case the errors are signed indicating 
in which direction the result changes if the 
underlying physics variable is changed in the 
direction indicated in the table. 
More details of the procedure and of all systematic errors are given in
\cite{bib-ctodstar}. 
\begin{table}
\begin{center}
\begin{tabular}{|l|l|c|}
\hline
Error source            &Variation &  Error \\
\hline\hline
\multicolumn{2}{|l|}{Total uncorrelated}               & $\pm$0.021 \\
\hline\hline
B mixing                & $\chi_{\rm eff}: + 11\%$ & $-$0.002 \\
Fragmentation modelling & $\langle x_E \rangle_{\rm B}: +0.008, 
                           \langle x_E \rangle_{\rm D}: +0.009$
                                                   & $+$0.004 \\
Gluon splitting         & $\bar n _{{\rm g}\to\ccbar}: +21 \%$ & $-$0.002 \\
\hline
\hline
\multicolumn{2}{|l|}{Total}                    & $\pm$ 0.022\\
\hline
\end{tabular}
\smcap{\label{tab-dstar}List of the systematic errors on the 
charm fraction \protect$\fc$ in the $\PDsp$ sample. The top 
part of the table contains that part of the error which is uncorrelated
with  the systematic error associated to the 
reconstruction of leptons in the $\PDsp$ sample. 
% the bottom part the errors which are correlated. 
The signs given for the errors in the 
lower part indicate the direction in which the 
result changes for a change of the relevant 
variable by the amount and direction indicated in the middle column.}
\end{center}
\end{table}
  
\section{The \boldmath{$\PDsp \ell^-$} Sample} 
\label{sec-leptag}
The $\PDsp \ell^-$ sample is found  by searching the 
hemisphere opposite the identified $\PDsp$ meson for a lepton with a 
charge opposite to that of the $\PDsp$ candidate.
Electrons are identified using a neural network technique~\cite{bib-OPALANN}. 
The network used in this part of 
the analysis
%\footnote{The network used here is not the same as is
%used for the electron identification in the semileptonic $\PDsp$
%decay channel, though it is very similar.} 
is slightly simplified compared to the one used in~\cite{bib-OPALANN}, 
using only 6 inputs, 
8 nodes in one hidden layer, and one 
output. The input variables are
\begin{itemize}
\item the difference between the measured specific energy loss, $\dEdx$,  
and that expected for an electron, divided by 
its expected uncertainty;
\item the experimental uncertainty on $\dEdx$; 
\item $E/p$, the energy of the 
electromagnetic cluster associated with the track inside a cone with a  
half opening angle of $30~\mathrm{mrad}$, divided 
by the measured track momentum; 
\item the number of  electromagnetic blocks in the cluster;
\item the momentum of the track;
\item the polar angle, $|\cos \theta|$, of the track.
\end{itemize}
All variables are well modelled in the Monte Carlo simulation, thus ensuring a reliable calculation of 
the selection efficiency. 

Muons are identified based on the $\chi^2$ of the 
matching between track segments in the central tracking chambers and in the muon 
chambers~\cite{bib-OPALleptonID}. 
In addition the specific energy loss, $\dEdx$, has to be compatible 
with that expected for a muon at the measured momentum. 

To reduce systematic uncertainties, electrons are reconstructed only 
in the central part of the OPAL detector, $|\cos\theta| < 0.715$, 
while muons are required to satisfy $|\cos\theta| < 0.9$. 
To increase the purity of the electron 
and muon samples, candidate tracks must have momenta
greater than $2~\GeVc$. Events from bottom decays are 
suppressed by selecting only candidates with $p_t<1.2 ~\GeVc$ for both electrons 
and muons, where the transverse momentum, $p_t$, is measured with respect to 
the axis of the jet containing the lepton candidate, 
including the lepton candidate itself in the jet-axis calculation. 
After all cuts, a total of $661$ electron and $1045$ 
muon candidates are selected. 
% opposite to a $\PDsp$ meson having opposite charge. 

\begin{figure}[p]
\begin{center}
\mbox{\epsfig{figure=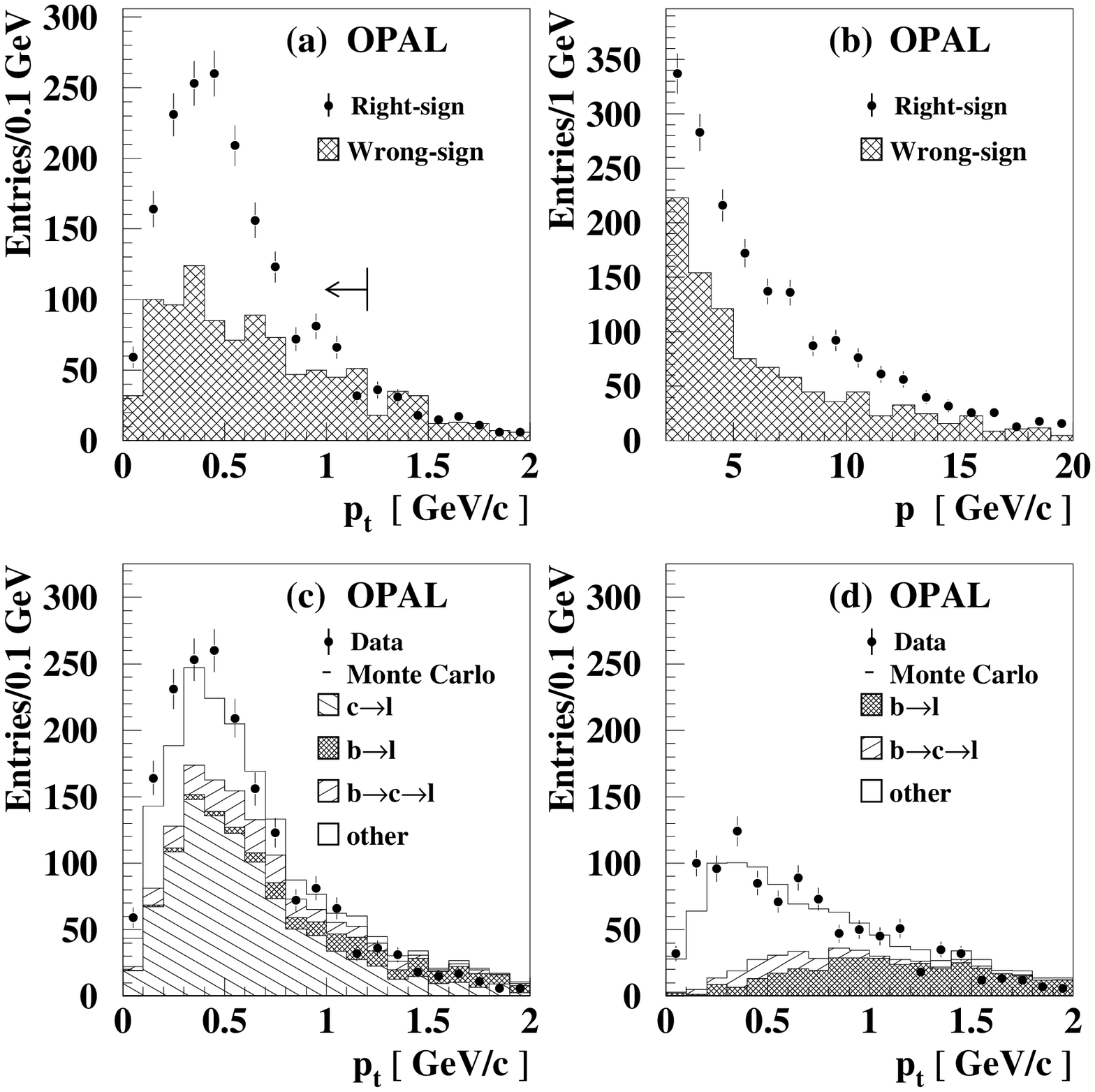,height=17cm,bbllx=2.5cm,bblly=2cm,bburx=18.5cm,bbury=22cm}}
\end{center}
\smcap{\label{fig-pt} Transverse momentum spectrum (a) and 
momentum spectrum (b) of the selected lepton candidates. 
The arrow in (a) indicates the position of the $p_t$ cut.
The hatched distribution is the background estimated using the 
wrong sign event sample. 
Composition of the $p_t$ spectrum in the Monte Carlo for 
right sign (c), and for wrong sign events (d).}
\end{figure}

The efficiency to select a lepton is calculated using Monte Carlo simulations. 
It is calculated from events 
where a $\PDsp$ meson is reconstructed in one hemisphere, 
and a lepton in the other, so that possible correlations between both hemispheres 
are taken into account. 
The $\epsilon_{\ctol}$ efficiencies are found to be 
\begin{eqnarray} 
\epsilon_{\ctoe} = 0.302 \pm 0.007 ~~~~{\rm and }~~~~  
\epsilon_{\ctom} = 0.433 \pm 0.008,
\end{eqnarray} 
where the errors are due to the finite Monte Carlo statistics.
%These efficiencies include the effects from the momentum, 
%from the transverse momentum and from the $|\cos\theta|$ cuts. 
A list of all  
efficiencies, including those for leptons in bottom events, 
is given in table~\ref{tab-eff}. The overall difference in 
the efficiencies for muons and electrons is mostly 
due to  the larger range of $\cos\theta$ used for the 
muons. The ratio $\epsilon_{\Pb\to\ell}/ \epsilon_{\Pc\to\ell}$ 
is larger for electrons than 
it is for muons, because the electron 
identification algorithm depends more 
strongly on $p_t$ than the muon identification does, 
the former being more efficient at large $p_t$. 
Since $\Pb\to\ell$ events have on average a 
larger $p_t$, electrons are found with larger 
efficiency in $\Pb\to\ell$ events.
%For $\Pb\to\ell$ events, however, the electron 
%identification efficiency is larger
%improves since 
%is relatively more efficient than the muon identification
%these events have, on 
%average, a larger $p_t$, such that the leptons 
%are well separated from the jet. The electron identification 
%for such events is significantly more efficient than 
%for electrons at low $p_t$, while the muon identification 
%does not significantly depend on $p_t$.
\begin{table}[tb]
\begin{center}
\begin{tabular}{|l|c@{$\ \pm \ $}c|c@{$\ \pm \ $}c|}
\hline
       & \multicolumn{4}{c|}{Efficiencies for }\\
Source & \multicolumn{2}{c|}{Electrons} & \multicolumn{2}{c|}{Muons}
\\
\hline\hline
$\ctol$    & 0.302 & 0.007 & 0.433 & 0.008 \\
$\btol$    & 0.305 & 0.015 & 0.305 & 0.015 \\
$\btoctol$ & 0.222 & 0.013 & 0.346 & 0.015 \\
$\btocbartol$& 0.213&0.031 & 0.315 & 0.036 \\
\hline
\end{tabular}
\smcap{\label{tab-eff}{Efficiencies to reconstruct an electron or 
a muon opposite a \protect$\PDsp$ meson separately for the 
different sources after applying all cuts. The errors quoted 
are purely statistical.}} 
%For bottom events the efficiency for 
%$\btol$ contains the contributions from $\btocbartol$ and 
%$\btotautol$. 
%Note that these efficiencies include the momentum 
%and the transverse momentum cuts.}
\end{center}
\end{table}
The $p_t$ and $p$ distributions of 
the selected candidates are shown in 
figures~\ref{fig-pt}a and \ref{fig-pt}b, respectively. 
The distributions of the wrong sign candidates are superimposed. 
%Here $p_t$ is the momentum of the lepton measured relative to the 
%direction of the jet containing the lepton, excluding the 
%lepton itself, and $p$ is the measured momentum of the lepton candidate 
%track. 

\subsection{Combinatorial Background Estimation}
% in the \boldmath$\PDsp \ell^-$ Sample
\label{sec-bgd}
Background in the $\PDsp \ell^-$ events is estimated from 
the data with the help of the wrong sign $\PDsp \ell^+$
sample. Subtracting the number of events found 
in the wrong sign sample from the number of events found in the 
right sign sample gives an estimate of the number 
of $\ctol$ decays, with only a small contribution remaining from 
background events (see equation~\ref{eq-ctol}).
The compositions of the right sign and 
of the wrong sign samples are shown in figures~\ref{fig-pt}c and 
\ref{fig-pt}d.

\begin{table}[tb]
\begin{center}
\begin{tabular}{|l|c@{$\ \pm \ $}cc|c@{$\ \pm \ $}c|c@{$\ \pm \ $}c|}  
\hline
Source        &  
\multicolumn{2}{|c}{Branching}  & & 
                                 \multicolumn{4}{c|}{Sample composition } \\   
              &
\multicolumn{2}{|c}{ratio} & ref. & 
                                \multicolumn{2}{c|}{Electrons} & 
                                \multicolumn{2}{c|}{Muons} \\
\hline\hline
$\ctol$      &\multicolumn{2}{c}{-}& & 
                                      $ 0.604$ & $0.019$ & $0.580$&$0.010$\\
$\btol$      & $0.1099$ & $0.0023$  & \cite{bib-PDG} & 
                                      $ 0.083$ & $0.011$ & $0.057$&$0.007$\\
$\btoctol $  & $0.0780$ & $0.0060$  & \cite{bib-PDG} &
                                      $ 0.135$ & $0.014$ & $0.098$&$0.010$\\
$\btocbartol$& $0.0130$ & $0.0050$  & \cite{bib-LEPNIM} & 
                                      $ 0.030$ & $0.006$ & $0.017$&$0.004$\\
$\btotautol$ & $0.0045$ & $0.0007$  & \cite{bib-PDG} & 
                                      $ 0.003$ & $0.002$ & $0.005$&$0.004$\\
$\btoJPsitol$& $0.0007$ & $0.0001$  & \cite{bib-PDG} & 
                                      $ 0.003$ & $0.002$ & $0.001$&$0.004$\\
Non-prompt   &\multicolumn{2}{c}{-}& & $ 0.063$ & $0.009$ & $0.089$&$0.008$\\
Hadrons $\ell$    &\multicolumn{2}{c}{-}& & $ 0.079$ & $0.009$ & $0.153$&$0.011$\\
\hline
\end{tabular}
\vspace{1cm}
\smcap{\label{tab-composition} Semileptonic branching ratios as given 
in \cite{bib-LEPNIM} and \cite{bib-PDG} and composition 
of the leptons in the \protect$\PDsp$ sample,
as found in the Monte Carlo. Note that the sample 
composition is given for information purposes only, and is not 
used in the actual analysis.}
\end{center} 
\end{table}

\begin{figure}[t]
\begin{center}
\mbox{\epsfig{figure=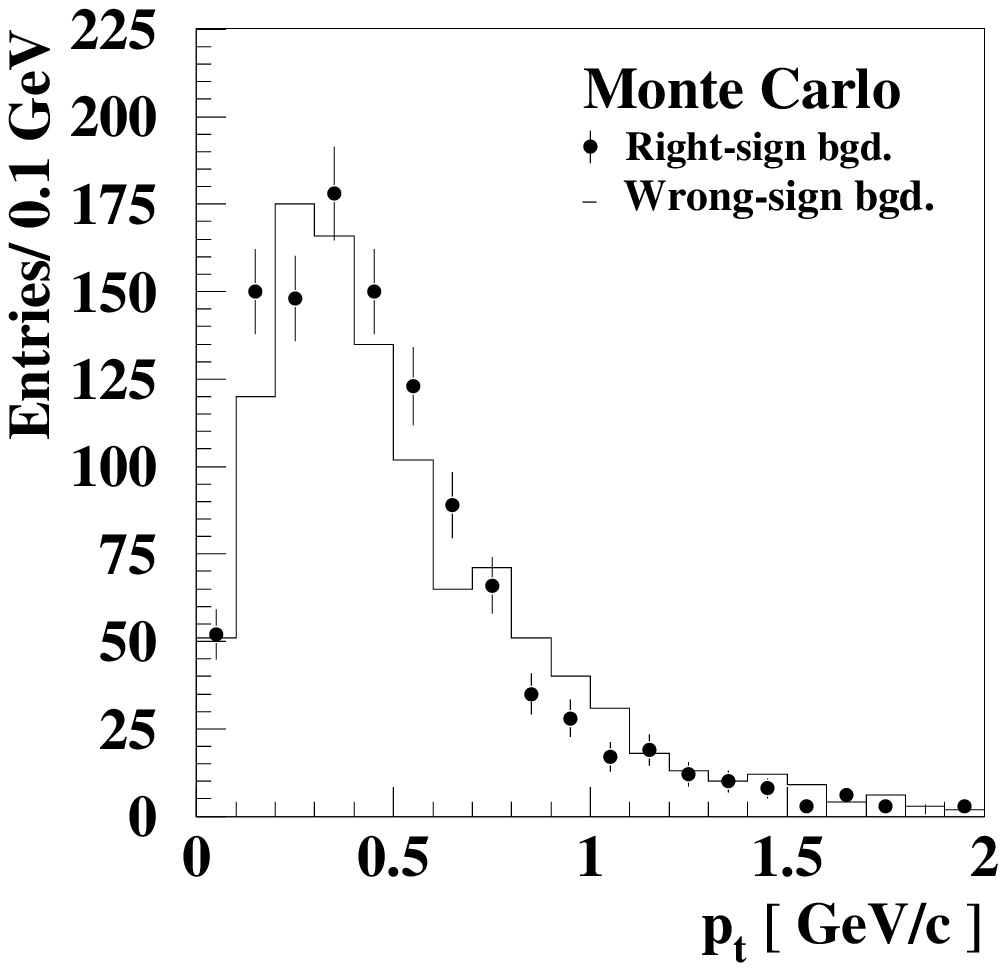,bbllx=5cm,bblly=5cm,bburx=16cm,bbury=13cm,height=7cm}}
\end{center}
\smcap{\label{fig-bgd}Comparison of the right sign combinatorial 
background (points with error bars) and the wrong sign combinatorial
background component (line histogram) in the simulation.}
\end{figure}

The subtraction of 
the combinatorial background relies on the assumption
that these events are equally distributed between the right 
and the wrong sign sample, namely that $N_{\rm bgd}^{+-}=N_{\rm bgd}^{++}$.
This subtraction procedure requires no explicit knowledge 
of the hadronic contamination in the lepton sample, 
since it is subtracted together with the wrong sign events. 
In figure~\ref{fig-bgd}, the shape of the $p_t$ distribution of 
the combinatorial background in the right 
sign sample, $N_{\rm bgd}^{+-}$,
is compared to the combinatorial component in the 
wrong sign sample, $N_{\rm bgd}^{++}$. 
Good agreement is observed for the fraction of events 
below the applied cut of $1.2~\GeV$
in the right and in the wrong sign combinatorial background.
The shapes are slightly different which is attributed to 
different contributions from bottom events to 
both samples. However since only the overall number 
of events is needed in this analysis the difference 
has a very small influence on the final result. 

%It is important to note that this difference has no impact on the analysis itself,
%since the shape distribution is not used. 

Monte Carlo studies show that the assumption 
$N_{\rm bgd}^{+-}=N_{\rm bgd}^{++}$ is not 
entirely correct, since a particular class of 
events, accounting for less than $10\%$ of the background, 
is found more often in the right sign sample than 
in the wrong sign sample.
These events consist of a partially reconstructed 
$\PDsp$ meson opposite a correctly identified lepton with the correct charge correlation.  
The number of such events found in the wrong sign sample amounts 
to only $55\%$ of the number of the same type of events 
found in the right sign sample. 
%Only $55\%$ of the number of events in the right sign sample are found in 
%the wrong sign sample. 
The total number of these events in the right sign sample has been 
measured in \cite{bib-ctodstar} from data. Relative 
to the combinatorial background, they account for 
$(8.5 \pm2.2)\%$ of the total right sign sample. The background 
subtracted sample is therefore corrected for the fraction of 
these events, 
%that are missing in the wrong sign sample, 
namely by 
$+45\%$ of the $(8.5\pm2.2)\%$. This corresponds to a 
background charge asymmetry of
$\Delta N_{\rm bgd}=+(43 \pm 11)$ events, where the error is dominated by 
the fraction of such events measured in the data. 
An additional modelling 
error of $50\%$ of this correction is applied,
as will be discussed in section~\ref{sec-sys}.
%distributions, even though the shapes are slightly different.
%The composition of 
%the right and the wrong sign samples is 
%illustrated in figures \ref{fig-pt}c and \ref{fig-pt}d. 

% \subsection{Flavour Composition of the \boldmath{$\PDsp\ell^-$} Sample}
\subsection{Estimation of the Bottom Background}
\label{sec-mixing}
The sample of tagged $\PDsp \ell^-$ events has a charm purity of about
$60\%$. Non-leptonic background 
accounts for $8\%$ of the electron candidates and $15\%$ of the muon candidates.
The rest consists of correctly identified leptons from a number of different
sources. The sample composition as determined from  the Monte Carlo simulations 
is shown in table~\ref{tab-composition}, and illustrated in 
figures \ref{fig-pt}c and \ref{fig-pt}d.
Since the charges of the $\PDs$ and the $\ell$ should be 
opposite in the $\PDsp\ell^-$ sample, any effect which influences
the charge correlation between the two 
hemispheres influences the flavour composition.
The most important of these is B-meson mixing.
If mixing has occurred in either hemisphere, the charge correlation 
between the primary quark and  the corresponding $\PDsp$ meson is changed.
The total probability in b-events that mixing has changed the charge 
correlation is given by 
\begin{eqnarray}
\chi_{\PDsp\ell^-} &=& \chi_{\PDsp}(1-\chi_{{\ell}}) +
                 \chi_{{\ell}}(1-\chi_{\PDsp})\ ,
\end{eqnarray}
where  $\chi_{\PDsp},\chi_{\ell }$ are the 
effective mixing parameters applicable to the $\PDsp$ and the lepton, 
respectively. 
These effective mixing parameters depend on the fractions of 
B$_{\Pd}^0$ and B$_{\Ps}^0$
mesons in the sample under consideration, 
and on the mixing in the B$_{\Pd}^0$ and the B$_{\Ps}^0$ system.
The average mixing in the B$_{\Pd}^0$ system is measured to be 
$\chi_{\Pd} = 0.175\pm 0.016$ \cite{bib-PDG}.
The LEP combined lower limit for $\PB_{\Ps}^0$
mixing given in~\cite{bib-PDG} corresponds
to a lower limit on $\chi_{\Ps}$ of 0.49 at $95\%$ confidence level. 
In this analysis, 
$\chi_{\Ps}$ is varied between 0.49 and the maximum value of $0.50$.

Most $\PDsp$ mesons in  $\PZz\to\bbbar$ events originate from decays of the 
B$_{\Pd}^0$ meson. In \cite{bib-mix2} this fraction has been 
determined to be $r_d^{\PDsp} = 0.81 ^{+0.05}_{-0.11}$. 
The fraction of $\PDsp$ that come from  B$_{\Ps}^0$ mesons 
has been estimated to be $r_s^{\PDsp} = 0.043 \pm 0.039$ 
\cite{bib-mix2}. The effective mixing in the hemisphere containing 
the $\PDsp$ meson is therefore 
\begin{equation}
    \chi_{\PDsp} = 
r_{\rm d}^{\PDsp} \cdot \chi_{\rm d} + r_{\rm s}^{\PDsp}\cdot
\chi_{\rm s} 
    = 0.163 ^{+0.025}_{-0.030}\nonumber \ .
\end{equation}
The fraction of leptons produced in decays of B$_{\Pd}^0$ and B$_{\Ps}^0$
mesons is determined from the fractions of weakly decaying 
B-hadrons in $\PZz\to\bbbar$ events by weighting with the 
lifetimes of the B-hadrons species~\cite{bib-PDG}. This is done  in 
order to correct for the different semileptonic branching ratios 
and leads to the values 
$r_d^{\ell} = 0.399 \pm 0.023 $ and $r_s^{\ell} = 0.118 ^{+0.019}_{-0.020}$,
respectively.  The effective mixing parameter is
\begin{equation}
    \chi_{\PLm}  = r_{\rm d}^{\ell} \cdot \chi_{\rm d} + 
                     r_{\rm s}^{\ell} \cdot \chi_{\rm s} 
   = 0.128 \pm 0.012 \ .\nonumber
\end{equation}
In addition, $\PDsp$ mesons with the wrong sign can be produced 
in bottom decays, where a $\Pac$ quark is produced in the decay of 
a virtual W. This can be expressed in terms of a mixing-like parameter 
$\zeta_{\rm D}$. As in \cite{bib-OPALD*}, a value of 
$\zeta_{\rm D}=0.025 \pm 0.025$ is used. The effective 
mixing parameter is then
\begin{equation}
\chi_{\rm eff} = \chi_{\PDsp\ell^-} (1-\zeta_{\rm D}) + 
                  \zeta_{\rm D} (1-\chi_{\PDsp\ell^-})\ ,
\end{equation} 
neglecting terms which are quadratic in either $\chi_{\PDsp\ell^-}$ 
or $\zeta_{\rm D}$.
The effective mixing parameter 
for the $\PDsp\ell^-$ sample is finally estimated to be
\begin{equation}
\label{eq-chieff} 
    \chi_{\rm eff} = 0.261 ^{+0.031}_{-0.034}.
\end{equation}   
In total, the contribution to the background from bottom events
is calculated according to equation~\ref{eq-B}, 
using the efficiencies listed in 
table~\ref{tab-eff}, the branching ratios given 
in table~\ref{tab-composition}, and the effective mixing 
parameter determined above. The total contribution amounts to $\Delta N_{\Pb}=-(53\pm7)$ events.

\section{Results}
\label{sec-results}
\begin{figure}[p]
\begin{center}
\mbox{\epsfig{figure=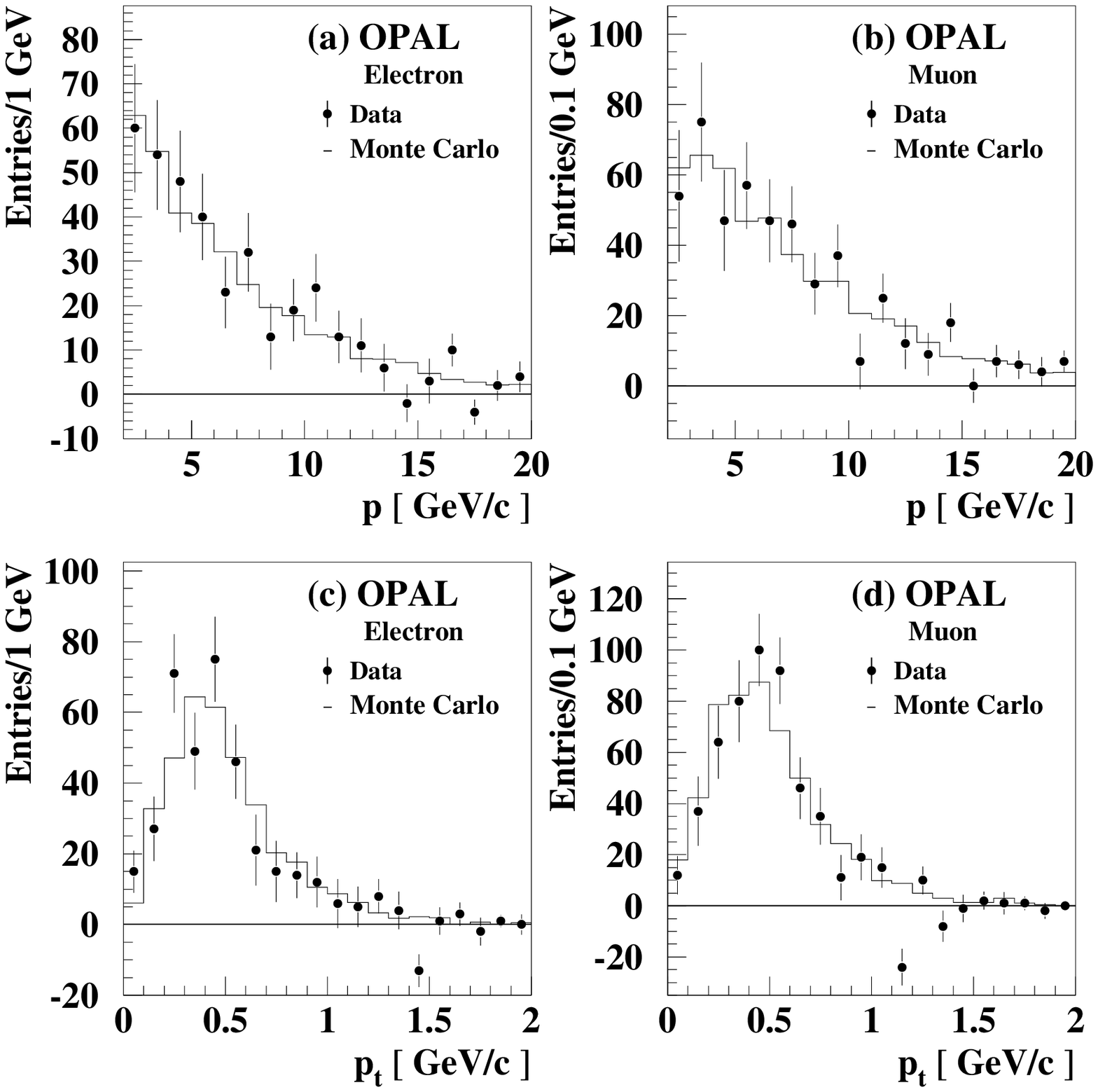,height=18cm,bbllx=3cm,bblly=0cm,bburx=15cm,bbury=20cm}}
\end{center}
\smcap{\label{fig-signal}Momentum spectra after background 
subtraction for electrons (a) and muons (b), and transverse momentum 
spectra for electrons (c) and muons (d). Points are data, 
the line histogram is the Monte Carlo prediction. Both data and Monte 
Carlo include the residual background contributions 
from bottom events and from the background charge asymmetry.}
\end{figure}

The number of $\PDsp \ell^-$ combinations in charm events 
is determined according to 
equation~\ref{eq-ctol}. The background 
subtracted momentum and transverse momentum spectra for 
electrons and muons are shown separately in figure \ref{fig-signal}.
The distributions are further corrected for the effects of mixing and 
for the charge asymmetry in the background, as
described in section~\ref{sec-bgd}.

%In figure \ref{fig-pt}, the right and wrong sign distributions are shown. 
%Figure \ref{fig-pt}c shows the 
%difference of both distributions together with the Monte Carlo prediction.
%Within errors, the overall agreement is good.  
The total number of leptons from charm hadron decays is 
$N_{\PDsp,\PEm}^{\Pc}  = 378 \pm 31$  and 
$N_{\PDsp,\PMm}^{\Pc} = 476\pm 40$, respectively. 
The quoted error is purely statistical.   
% \mbox{$N_{\PDsp,\PLm}^{\Pc} = 854 \pm 51$ $\PDsp\ell^-$} events. 
In table~\ref{tab-sample}, a summary of the selected events in each sample 
is shown.  
Combining  these measurements with the total number of selected $\PDsp$ mesons,
$N_{\PDsp} = 15784 \pm 99$, the
appropriate charm fraction, and the lepton efficiencies, the inclusive semileptonic branching 
ratios of charm hadrons in $\PZz\to\ccbar$ events are determined to be
$$ \fctoe = \nctoe \pm \estatctoe ~~~{\rm and} ~~~ 
   \fctom = \nctom \pm \estatctom\ .$$
Here, the quoted errors are only statistical. 
The semileptonic branching fraction of charm hadrons derived from 
these individual results is 
$$ \fctol = \nctol \pm \estatctol.$$

\begin{table}[t]
\begin{center}
\begin{tabular}{|l|c|c|}  
\hline
Sample            & $\PDs\PE$     & $\PDs\PM$   \\ \hline\hline
Right-sign        &   661          & 1045       \\ 
Wrong-sign        &   305          & 558        \\ 
\hline
$\NDLc$           &   $378\pm31$   & $476\pm40$  \\
\hline
\end{tabular}
\smcap{\label{tab-sample} Summary of selected events in each sample,
  and number of events after background subtraction.}
\end{center} 
\end{table}

\section{Systematic Errors}
\label{sec-sys}
In this section, the different sources of systematic errors are
discussed. 
A breakdown 
of all errors considered is summarised in table \ref{tab-syserror}. 
All errors in this section are given relative to the
inclusive $\fctol$ branching ratio, if not otherwise stated. 
  
\newpage
\begin{itemize} 
\item Modelling errors
\begin{itemize} 
\item $\ctol$ modelling:
The momentum spectrum of leptons in $\ctol$ decays 
is described by the ACCMM model. 
Using the range of parameters recommended in ~\cite{bib-LEPNIM} 
this corresponds to an error of 
$\;^{+5.6}_{-3.3}\%$ of the charm semileptonic branching ratio. The 
size of the error is largely dependent on the momentum cut 
used in the identification of electrons and muons. 
%The 
%stability of the analysis is checked by increasing the 
%cut to $3~\GeVc$ and repeating the analysis and the 
%determination of the systematic errors. Consistent results 
%are found, though with a slightly larger systematic error. 
%
\item $\btol$ modelling: 
The momentum distribution of the leptons in bottom decays 
influences the tagging efficiencies. Following the recommendations 
in~\cite{bib-LEPNIM} this has been studied by 
reweighting the lepton spectrum in the Monte Carlo simulation
to different theoretical models, with ranges of parameters 
chosen such that the experimental errors are covered. 
The ACCMM~\cite{bib-ACCM} model is used to obtain the central value, and 
the ISGW and the ISGW$^{**}$~\cite{bib-ISGW} models are used for 
the $\pm 1 \sigma$ variation around the central value,  
and the efficiencies are recalculated.  
The errors found are $0.1\%.$ 
\item $\btoctol$ modelling:
The ACCMM model is used to describe the momentum 
spectrum of cascade $\btoctol$ 
decays, as suggested in \cite{bib-LEPNIM}. 
Three different sets of parameters are proposed to 
cover the experimental uncertainties in the 
momentum spectrum.
The lepton efficiencies are recalculated.  
The errors for the final result are 
$\;^{+0.2}_{-0.1}\%$.
% by varying the model parameters in the 
%ranges suggested in \cite{bib-LEPNIM}.
%
\item Fragmentation modelling: 
The fragmentation 
parameters in the Monte Carlo simulation have been varied to change 
the mean scaled energy of weakly decaying bottom and charm hadrons around 
their experimental values of
$\langle x_{E}\rangle_{\rm B}=0.702\pm 0.008$ and 
$\langle x_{E}\rangle_{\rm D}=0.484\pm 0.009$
respectively~\cite{bib-LEPNIM}. 
This study is done using the Peterson fragmentation model~\cite{bib-PETERSON}. 
This results in an error of $\pm 0.6\%$.
In addition, the Peterson fragmentation model has been replaced by the 
Collins and Spiller fragmentation model~\cite{bib-collinsspiller}
and by the Kartvelishvili fragmentation model~\cite{bib-kart}.
The parameters for these models have been adjusted to the 
same mean scaled energy as for the Peterson function. 
The largest deviation between the different 
models is used as a systematic error.
Combined with the error using different parameters 
for the Peterson model, 
a total error of $0.9\% $ is determined. 
  \end{itemize}
\end{itemize}

\begin{itemize}
\item B-physics
\begin{itemize}

\item{B-meson mixing: 
The uncertainty due to mixing in the neutral B sector has been studied
by varying the effective mixing parameter (see equation~\ref{eq-chieff}) 
$\chi_{\rm eff}$ within its errors. An error of $0.8\% $ is found. }
\item Branching ratios: The dependence on the branching ratios
$\btol$ and  $\btoctol$ 
has been investigated by 
varying them within their experimental errors. 
Mean values and errors used are given in table~\ref{tab-composition}.
An error of $0.8 \%$ is found. A breakdown of the error into 
the different channels contributing is given in table~\ref{tab-syserror}. 
\end{itemize}
\end{itemize}

\begin{itemize}
\item Particle identification
\begin{itemize}
\item Electron identification: The efficiency to identify electrons
is calculated in the Monte Carlo. The two variables which mainly determine
the performance of the neural network are the specific energy loss, 
$\dEdx$, together with its error, 
and the ratio $E/p$. Both variables are compared 
between Monte Carlo and data using different samples 
of identified particles. The $\dEdx$ measurements are 
calibrated in data using samples of inclusive pions 
at low momenta and electrons from Bhabha events at $45~\GeVc$. 
The quality of the calibration is checked with a number 
of control samples, mostly pions from $\PKs$ decays and electrons 
from photon conversions. The deviation between the 
mean $\dEdx$ measured for these samples in the data, 
and the mean $\dEdx$ in the Monte Carlo, is below $5\%$. 
Similarly, the resolution of $\dEdx$ is studied in these 
samples, and is found to be described in the Monte 
Carlo to better than $8\%$. The total error from these two 
effects is found by varying both simultaneously, and is 
%$^{+1.9}_{-2.5}\%$. 
$\pm 2.5\%$
Note that for this analysis, no explicit knowledge of the hadronic 
background in the sample of lepton candidates is needed, 
since it is subtracted using the wrong sign sample. 

A similar study has been performed for the 
next most significant input variable, $E/p$. 
The $E/p$ resolution in the Monte Carlo is
about $10\%$ worse than in the data. The Monte Carlo has 
been reweighted to the data, and the full difference is 
used as an estimate of the error, resulting 
in a variation of the efficiency of 
%$\;^{+1.1}_{-2.7}\%$.
$\pm 2.7\%$.

No significant contributions to the error are found from 
the other input variables of the network. The error related 
to them is estimated from the statistical precision 
of these tests, which is less than $1\%$ of the efficiency. 
In total, an error of $\pm 4\%$ is assigned to this source. 
\item Muon identification: The systematic error of the 
muon identification efficiency is evaluated using a 
method similar to that 
described in \cite{bib-OPALleptonID}.  
The muon detection efficiency is compared between data and Monte
Carlo using various control samples, namely $\PZz\to\mu^+\mu^-$
events, and muons reconstructed in jets.
Without using $\dEdx$ information, an error of $\pm 2\%$ is found. 
The influence of the $\dEdx$ selection 
cut on the muon ID is studied in the same way as described 
for electrons. The mean $\dEdx$ for muons 
in $\PZz\to\mu^+\mu^-$ events is observed to be shifted by 
approximately $15\%$ of the resolution in $\dEdx$ with respect to
the theoretically expected value. A very similar shift is observed 
in the Monte Carlo, both for muon pairs and for muons 
identified inclusively in jets.
An error of $5\%$ is used.
% as an estimate of the systematic error. 

The $\dEdx$ resolution is studied in the data, 
and is found to be modeled by the Monte Carlo to better than $5\%$. 
The final error assigned to the efficiency of muon 
identification is $\pm 3.0 \%$.
\end{itemize}
\end{itemize}

\begin{itemize}
\item Internal sources
\begin{itemize}
\item{Flavour separation:  
The errors of the flavour composition on the $\PDsp$ sample  
estimated in \cite{bib-ctodstar} are used to calculate the 
corresponding error of the semileptonic branching fractions. 
A breakdown of the total error into sources correlated 
and uncorrelated with the reconstruction of leptons in 
the $\PDsp$ sample is given in table~\ref{tab-dstar}, 
and is taken into account in calculating the final error. The 
uncorrelated error is $2.8\%$; the total correlated error is  $0.9\%$.}
\item Background charge asymmetry:
The correction applied to the background-subtracted sample 
of $\PDsp\ell^-$ events is calculated based on the 
measured fraction of events contributing, and on 
the charge asymmetry, which comes from Monte Carlo simulation. 
For the former, the statistical error of the measurement is 
used as a systematic uncertainty, translating into an 
error of $1.1\%$. The Monte Carlo 
prediction of the charge asymmetry is conservatively varied by 
$\pm 50\%$ of its value. The final error from this 
is $2.5\%$.
% lepton_charge.kumac
%
\item Background estimation:
The background in the $\PDsp\ell^-$ sample is estimated 
from the wrong sign sample. 
The number of combinatorial background events, corrected 
for mixing and for the effects of the background charge 
asymmetry, is compared with the expected number of 
combinatorial background events in the Monte Carlo simulation. 
Within the statistical precision of this test 
good agreement is found. The statistical error of 
this test is used as a systematic uncertainty, resulting 
in an error of $1.8\%$. The influence of the 
$p_t$ cut on the background is studied by comparing the 
shapes of the background between data and Monte Carlo. 
The data spectrum is reweighted to the Monte Carlo 
one, and the number of background events is 
recalculated. The resulting difference is used as a 
systematic error of $0.3\%$.
The final error assigned is $1.8\%$.
%Possible modelling 
%problems are studied by reweighting each component to 
%its true value, and evaluating the effect on the 
%final number of background events. The error 
%is estimated by adding the individual errors 
%in quadrature, and is $0.3\%$. To allow for possible 
%additional modelling problems the statistical precision of 
%the first test of $3\%$ is also added as an error, 
%resulting in a total error of $1.8\%$.
%
% clep/rq-wq.kumac 
%
%
\item{Detector modelling: The influence of the detector 
resolution on the tagging variables is studied in Monte Carlo 
simulations by varying the resolutions in the central tracking 
detectors 
by $\pm 10\%$ relative to the values that optimally describe the
data. The analysis is repeated and the efficiencies are recalculated. 
The error is $1.1\%$.

The calculation of the efficiencies relies on the 
correct modelling of the detector acceptances, in particular 
in $\cos\theta$. This has been tested by reweighting
the $\cos\theta$ distribution of $\PDsp$ candidates as 
found in the Monte Carlo simulation to that reconstructed from data, 
and repeating the analysis. This changes the result by $0.3\%$, 
which is used as a systematic error. 
The total the error due to detector modelling  is $ 1.2 \%$.}
\item Gluon splitting: Gluon splitting into a pair 
of heavy quarks can produce $\PDsp$ 
mesons which might contribute to the 
sample of selected events. This contribution
is found to be $(1.1\pm0.4)\%$. It is based 
on the OPAL measurement of gluon splitting~\cite{bib-OPALgcc} and 
Monte Carlo simulation to determine the selection efficiency.  
The total number of $\PDsp$ mesons is corrected for this effect.
The uncertainty of 
this number is used as a systematic error of $0.4\%$.
Similarly, leptons can be produced in gluon splitting events.
The contribution to the sample is found to be 
$(0.2\pm0.1)\%$, which results in an error of $0.1\%$.
According to these studies the total systematic uncertainty is $0.4\%$.
\end{itemize}
\end{itemize}

\begin{itemize}
\item Monte Carlo statistics
\begin{itemize}
\item{Monte Carlo statistics: The efficiencies to identify 
a lepton in the $\PDsp$ sample are calculated from the 
Monte Carlo with limited statistical precision. The error from 
this source amounts to $1.5\%$. }
%\item Hadronic background: Around $6\%$ of the sample 
%of tagged leptons are hadrons, which were misidentified. In~\cite{bib-OPALRb},
%the error of the mistagging rate has 
%been determined to be $9.3\%$ in the electron sample, 
%and $9.0\%$ in the muon sample. This translates into an 
%error of $0.6\%$ of the bottom purity.
\end{itemize} 
\end{itemize} 
\begin{table}[p]
\begin{center}
\begin{tabular}{|l|c|c|c|}  
\hline
Source \rule[-1.9mm]{0mm}{6mm}& ${\fctoe}$ & ${\fctom}$ & ${\fctol}$ \\   
\hline\hline
Modelling &\multicolumn{3}{c|}{} \\
\hline
$\ctol$ model\rule{0mm}{5mm}
            & $^{+0.0057}_{-0.0034}$ 
                                       & $^{+0.0050}_{-0.0030} $
                                                 &$^{+0.0053}_{-0.0031}$ \\
$\btol$ model              &   0.0001  & 0.0001  & 0.0001\\
$\btoctol$ model           &   0.0002  & 0.0002  & 0.0002\\  
Fragmentation modelling    &   0.0010  & 0.0008   & 0.0009\\
\hline
Total modelling \rule[-1.7mm]{0mm}{6mm}
                           &   $^{+0.0058}_{-0.0035}$
                                       & $^{+0.0051}_{-0.0032}$
                                                 &$^{+0.0054}_{-0.0032}$ \\
\hline\hline
B physics     &\multicolumn{3}{c|}{}\\
\hline
B-meson mixing             &  0.0009  & 0.0008  & 0.0008\\
$\fbtol$                   &  0.0003  & 0.0002  & 0.0003\\
$\fbtoctol$                &  0.0007  & 0.0006  & 0.0006\\
$\fbtocbartol$             &  0.0005  & 0.0004  & 0.0005\\
$\fbtotautol$              &  0.0001  & 0.0001  & 0.0001\\
$\fbtoJPsitol$             &$<0.0001$ &$<0.0001$&$<0.0001$\\
\hline
Total B physics    
\rule[-1.7mm]{0mm}{6mm}     & 0.0013   & 0.0011  &0.0012 \\
\hline\hline
Particle ID &\multicolumn{3}{c|}{}\\
\hline
Electron identification    
%                           & $^{+0.0025}_{-0.0039}$ 
%                                       & -       
%                                                 &$^{+0.0010}_{-0.0016}$\\
                           &  0.0041    & -      &  0.0017\\
Muon  identification       &   -       & 0.0027  & 0.0015\\
\hline\hline
Internal sources &\multicolumn{3}{c|}{} \\
\hline
Flavour separation (uncorr.)&   0.0028  & 0.0024  & 0.0026\\
Background charge asymmetry &   0.0026  & 0.0023  & 0.0024\\
Background estimator        &   0.0019  & 0.0016  & 0.0017\\
Detector modelling           &   0.0012  & 0.0011  & 0.0011\\
%Branching ratios           &   0.0006  & 0.0005  & 0.0006\\
Gluon splitting             &   0.0005  & 0.0004  & 0.0004\\
\hline\hline
Monte Carlo statistics      &   0.0024  & 0.0016  & 0.0015 \\
\hline\hline
 Total error  \rule[-2.0mm]{0mm}{6mm}
                            & $^{+0.0088}_{-0.0075}$
                                       &$^{+0.0072}_{-0.0060}$
                                                 &$^{+0.0074}_{-0.0060}$\\
\hline
\end{tabular}
\vspace{1cm}
\smcap{\label{tab-syserror} List of systematic errors contributing
to \protect$ {\fctoe}$, \protect$\fctom$ and \protect$\fctol$. 
A detailed explanation 
of the different errors can be found in the text.}
\end{center} 
\end{table}
A complete list of systematic errors is presented in table~\ref{tab-syserror}
for $\fctoe$, $\fctom$, and $\fctol$. Except for the 
error from Monte Carlo statistics and the lepton identification 
errors, all errors from a given source are assumed 
to be fully correlated between the electron and the 
muon results. 

To check the stability of the results, the analysis is repeated
with different selection cuts for the leptons. Consistent 
results are found if the momentum cut is raised from 
$2\;\GeVc$  to $3\;\GeVc$ both for
electrons and muons, if the transverse momentum cut is
removed, 
if the muon selection is repeated using muons in the central 
part of the detector only, and
if the muon selection is done without using the 
$\dEdx$ selection cut.
%
%The stability of the analysis is checked by increasing the 
%cut to $3~\GeVc$ and repeating the analysis and the 
%determination of the systematic errors. Consistent results 
%are found, though with a slightly larger systematic error.
%

\section{Conclusions}
\label{sec-conclude}
A measurement of the inclusive charm hadron semileptonic 
branching fractions in $\PZz\to\ccbar$ events, 
$\fctoe$ and $\fctom$, has been presented.
The identification of $\PZz\to\ccbar$ events is 
based on the reconstruction of $\PDsp$ mesons. The semileptonic 
branching ratios are measured by reconstructing leptons  
in the charm-tagged sample and are found to be
$$ \fctoe = \nctoe \pm \estatctoe \esysctoe ~~~{\rm and} ~~~ 
   \fctom = \nctom \pm \estatctom \esysctom\ , $$
where the first error is in each case statistical and the second systematic. 
Combining the two measurements while taking correlations into account,
%the inclusive semileptonic branching fraction of charm hadrons is
%derived to be
gives
$$ \fctol = \nctol \pm \estatctol \esysctol \ . $$
%The largest contribution to the systematic error comes from 
%the modelling of the momentum spectrum of $\ctol$ decays, 
%amounting to $^{+0.005}_{-0.003}$. 
This result agrees well and is competitive with the most recent 
published measurement at  lower energies of 
$\fctol = 0.095 \pm 0.009$ \cite{bib-cl1}. 

%-----------------------------------------------------------------------
\section*{Acknowledgements}
%-----------------------------------------------------------------------

We particularly wish to thank the SL Division for the efficient operation
of the LEP accelerator at all energies
 and for their continuing close cooperation with
our experimental group.  We thank our colleagues from CEA, DAPNIA/SPP,
CE-Saclay for their efforts over the years on the time-of-flight and trigger
systems which we continue to use.  In addition to the support staff at our own
institutions we are pleased to acknowledge the  \\
Department of Energy, USA, \\
National Science Foundation, USA, \\
Particle Physics and Astronomy Research Council, UK, \\
Natural Sciences and Engineering Research Council, Canada, \\
Israel Science Foundation, administered by the Israel
Academy of Science and Humanities, \\
Minerva Gesellschaft, \\
Benoziyo Center for High Energy Physics,\\
Japanese Ministry of Education, Science and Culture (the
Monbusho) and a grant under the Monbusho International
Science Research Program,\\
Japanese Society for the Promotion of Science (JSPS),\\
German Israeli Bi-national Science Foundation (GIF), \\
Bundesministerium f{\"u}r Bildung, Wissenschaft,
Forschung und Technologie, Germany, \\
National Research Council of Canada, \\
Research Corporation, USA,\\
Hungarian Foundation for Scientific Research, OTKA T-016660, 
T023793 and OTKA F-023259.\\

\end{document}